\documentclass{jpsj-suppl}
\usepackage{txfonts} 
\usepackage{wrapfig}
\def \dru{day$^{-1}$kg$^{-1}$keV$^{-1}$}

\title{High purity NaI(Tl) scintillator to search for dark matter}

\author{Ken-Ichi \textsc{Fushimi}$^{1}$,  Hiroyasu \textsc{Ejiri}$^{2}$, Ryuta \textsc{Hazama}$^{3}$,
Haruo \textsc{Ikeda}$^{4}$, Kunio \textsc{Inoue}$^{4}$, Kyoshiro \textsc{Imagawa}$^{5}$,
Gakuji \textsc{Kanzaki}$^{6}$, Alexandre \textsc{Kozlov}$^{7}$, Reiko \textsc{Orito}$^{1}$, 
Tatsushi \textsc{Shima}$^{2}$, Yasuhiro \textsc{Takemoto}$^{7}$, Yuri \textsc{Teraoka}$^{4}$,
Saori \textsc{Umehara}$^{2}$ and Sei \textsc{Yoshida}$^{8}$}

\inst{$^{1}$Institute of Socio- Arts and Science, Tokushima University, 1-1 Minami Josanjimacho,
Tokushima city, Tokushima 770-8502, Japan \\
$^{2}$Research Center for Nuclear Physics, Osaka University, 1-10 Mihogaoka, Ibaraki city,
Osaka 567-0042, Japan\\
$^{3}$Graduate School and Faculty of Human Environment, Osaka Sangyo University, 3-1-1 Nakagaito Daito city, Osaka 574-8530, Japan\\
$^{4}$Research Center for Neutrino Science, Tohoku University, 6-3 Aramaki Aza Aoba 
Aoba ward Sendai city, Miyagi 980-8578, Japan\\
$^{5}$I.S.C. Lab., 7-7-20 Saito Asagi Ibaraki City, Osaka 567-0085, Japan\\
$^{6}$Graduate School of Integrated Arts and Sciences, Tokushima University, 1-1 Minamijosanjimacho Tokushima city, Tokushima 770-8502, Japan\\
$^{7}$Kavli Institute for the Physics and Mathematics of the Universe (WPI),
The University of Tokyo Institutes for Advanced Study, The University of 
Tokyo, 5-1-5 Kashiwanoha Kashiwa city, Chiba 277-8583, Japan\\
$^{8}$Department of Physics, Osaka University, 1-1 Machikaneyama
Toyonaka city,  Osaka 560-0043, Japan}

\email{kfushimi@tokushima-u.ac.jp}

\recdate{\today}

\abst{
A high purity and large volume NaI(Tl) scintillator was developed to search for cosmic dark matter. 
The required densities of radioactive impurities (RIs) such as U-chain, 
Th-chain are less than a few ppt to establish high sensitivity to dark matter. 
The impurity of RIs were effectively reduced by selecting raw materials of crucible and by performing chemical reduction of lead ion in NaI raw powder. 
The impurity of $^{226}$Ra was reduced less than 100 $\mu$Bq/kg in NaI(Tl) crystal.
It should be remarked that the impurity of $^{210}$Pb, which is difficult to reduce,
is effectively reduced by chemical processing of NaI raw powder down to less than 30 $\mu$Bq/kg.
The expected sensitivity to cosmic dark matter by using 250 kg of the high purity and large 
volume NaI(Tl) scintillator (PICO-LON; Pure Inorganic Crystal Observatory for LOw-background
Neutr(al)ino) is $7\times10^{-45}$ cm$^{2}$ for 50 GeV$/c^{2}$ WIMPs.
}

\kword{cosmic dark matter, inorganic crystal scintillator}

\begin{document}
\maketitle

\section{Introduction}

Cosmic dark matter is one of the most important problem in the fields of cosmology and 
particle physics.
There are many evidences for cosmic dark matter existence in the Universe and our galaxy
\cite{WMAP, Planck, COSMOS}.
The fraction of the energy density of cosmic dark matter is about 23 \% in the Universe, 
on the other hand, the fraction of atoms is only 5 \%\cite{WMAP, Planck}.
The dark matter must be the unknown particles which is proposed by the beyond standard model of 
the elementary particle physics.
Many models of the particle theories are proposed to solve not only the dark matter problem but also 
CP violation, hierarchy problem and so on \cite{DMReview}.
One of the most promising candidates for the dark matter particle is a sort of heavy elementary
particles which interact with the matter by weak interaction and gravity,
so called WIMPs (Weakly Interacting Massive Particles).

The dark matter fraction in the galaxy is more than 90 \% \cite{density}. 
The expected density of dark matter in the vicinity of our solar system is $0.3\sim0.5$ GeVcm$^{-3}$
\cite{density}.
The velocity of WIMPs is assumed to be isotropic in direction and Maxwellian distribution with 
a root mean square velocity of 230 km/sec.
The revolution of the earth around the sun and the motion of the solar system in the galaxy results
the annual modulation of the energy spectrum of the WIMPs-nucleus elastic scattering
\cite{Drukier, Freese, Miuchi}. 
The annual modulating signal gives a clear evidence for the detection of WIMPs.
The net velocity of the sun with respect to the galactic rest frame is 
$v_{\odot}=232\pm20$ km/sec.
The velocity of the Earth's revolution is 30 km/sec and the orbit is inclined with $\theta=30^{\circ}$
($\sin\theta=0.5$). Thus the velocity of the Earth with respect to the galactic rest frame has 
its maximum in the beginning with its velocity of $v_{max}=247$ km/sec and its minimum 
in the beginning of December with its velocity of $v_{min}=217$ km/sec.
Consequently, the event rate of scattering between WIMPs and atomic nuclei
above the energy threshold has its maximum in June and its minimum in December.

Many experimental efforts to search for WIMPs were
done by large volume detectors.
NaI(Tl) scintillator was applied by many groups; DAMA/LIBRA\cite{DAMA2013_EPJ}, 
ELEGANT V\cite{ELEVNaI}, PICO-LON\cite{FushimiKEK}, ANaIS\cite{ANaIS_TAUP2015},
KIMS\cite{KIMS_NaI}, DM-ICE\cite{DM-ICE2014} and SABRE\cite{SABRE_TAUP2015}.
The experiments using liquid Xenon scintillator gave more stringent limits than the NaI(Tl) 
experiments\cite{LUX_1st, XENON100, XMASS_LightWIMPs}.
Recently, XMASS groups excluded significant annual modulation in the region where
DAMA/LIBRA reported\cite{XMASS_annual}.
Some experiments using Ge and Si semiconductor gave significant annual modulating signals
\cite{CDMS2013, EDELWEISS}.
Note that all the significant modulating signals do not agree with each other.

The DAMA/LIBRA group reported a significant modulating signal\cite{DAMA2013_EPJ}.
The total mass of the NaI(Tl) was 100 kg for the first 6 years and upgraded mass of 250 kg for 
the last 7 years.
They continued the low background and low energy threshold measurement to search for significant 
annual modulation by WIMPs.
The change of the residual event rate between 2 keV and 6 keV was well fitted by a cosine 
function with the period $T=\frac{2\pi}{\omega}=1$ yr and a phase $t_{0}=152.5$ day (June 2nd)
as expected the annual modulation signature.

Their data have given positive model independent evidence for the presence of dark matter particles
with high confidence level \cite{DAMA2013_EPJ}.
However, the groups other than DAMA/LIBRA which are trying to use NaI(Tl) scintillator
have not given enough sensitivity to check 
the annual modulating signal because of the radioactive contamination.

The background rate of the large volume NaI(Tl) used by DAMA/LIBRA was 
1 \dru \ above 1 keV energy threshold.
The low background measurement by the large volume NaI was established by developing the highly 
radiopure NaI(Tl) crystal.
The impurities of RIs in their NaI(Tl) crystal was listed in Table \ref{tb:RI}.
\begin{table}[ht]
\caption{The RI concentration in NaI(Tl) crystals developed by other groups.}
\label{tb:RI}
\begin{tabular}{l|rrrrrr} \hline
RI & DAMA/LIBRA\cite{DAMA_NIM} & ELEGANT V\cite{ELEVNaI}  
 & ANaIS\cite{ANaIS_TAUP2015} & KIMS\cite{KIMS_NaI,KIMS_TAUP2015} & 
DM-ICE\cite{DM-ICE2014}  \\ \hline
Th chain(ppt) & 0.5$\sim$7.5 & 5.8$\pm$0.3 & 0.8$\pm$0.3 & 0.5$\pm$0.3 & 10 \\
$^{226}$Ra($\mu$Bq/kg) & 21.7$\pm$1.1 & 281$\pm$6 & 10$\pm$2 & $<1$ & 900 \\
$^{210}$Pb($\mu$Bq/kg) & 24.2$\pm$1.6 & 6910$\pm$200 & 600$\sim$700 & 470$\pm$10 & 1500 \\
$^{nat}$K(ppb) & $<20$ & $<200$ & 35$\sim$40 & 40$\sim$50 & 56 \\ \hline
\end{tabular}
\end{table}

\section{Outline of the PICO-LON project to dark matter search}
PICO-LON (Pure Inorganic Crystal Observatory for LOw-background Neutr(al)ino) aims at 
finding the signal by WIMPs by using highly radiopure NaI(Tl) scintillator.
NaI(Tl) scintillator is suitable for WIMPs detection in the region of $10^{-45}$ cm$^{2}$
because it is sensitive to both types of 
interactions, spin-independent (SI) and spin-dependent (SD).
The cross section of SI is proportional to the square of the mass number of the target nucleus.
The cross section of WIMPs-$^{127}$I scattering is enhanced by a factor of $A^{2}=16129$.
The cross section of SD depends of the finite value of nuclear spin $J$.\ 
Both $^{23}$Na and $^{127}$I have finite spins $J_{Na}=3/2$ and $J_{I}=5/2$.\ 
Both the natural abundances of $^{23}$Na and $^{127}$I are as large as 100\%.

The projected PICO-LON detector is an array of 42 modules of 12.70 cm$\phi\times$12.70cm
cylindrical 
crystal of NaI(Tl) and the total mass of the NaI(Tl) is 248 kg.
The total mass of the PICO-LON detector is large enough to check the annual modulating signal 
which is reported by DAMA/LIBRA\@.
The NaI(Tl) crystal is viewed by a low background photomultiplier tube (PMT) whose diameter 
is 10.16 cm through a high purity quartz light guide.
The NaI(Tl) and light guide are encapsulated in a copper container to prevent deliquesce.
The continuous low background measurement will be performed for several years to 
investigate the WIMPs whose cross section down to $10^{-43}$ cm$^{2}$.

The goal of the radioactive impurities was set to enable to find a candidate for the 
annual modulation signal of WIMPs which is expected to be  less than $10^{-2}$ \dru\@.
The radioactive impurity in NaI(Tl) crystal is required to give the sufficient small number of 
background rate at the energy threshold. 
The goal of the background is set to be less than 1 \dru\ which is the same result 
as DAMA/LIBRA\@.
If the background rate is more than 1 \dru\ , one can not study the modulation of 
$10^{-2}$ \dru \@.
The contamination in NaI(Tl) is needed to be less than a few ppt for U and Th chain, 
a few tens $\mu$Bq/kg for $^{210}$Pb and a few tens ppb for $^{nat}$K to perform the effective 
investigation of WIMPs existence.

\section{Purification of NaI(Tl) crystal}
\begin{wrapfigure}{r}{0.45\textwidth}
\centering
\includegraphics[bb=0 0 567 384 ,width=0.9\linewidth]{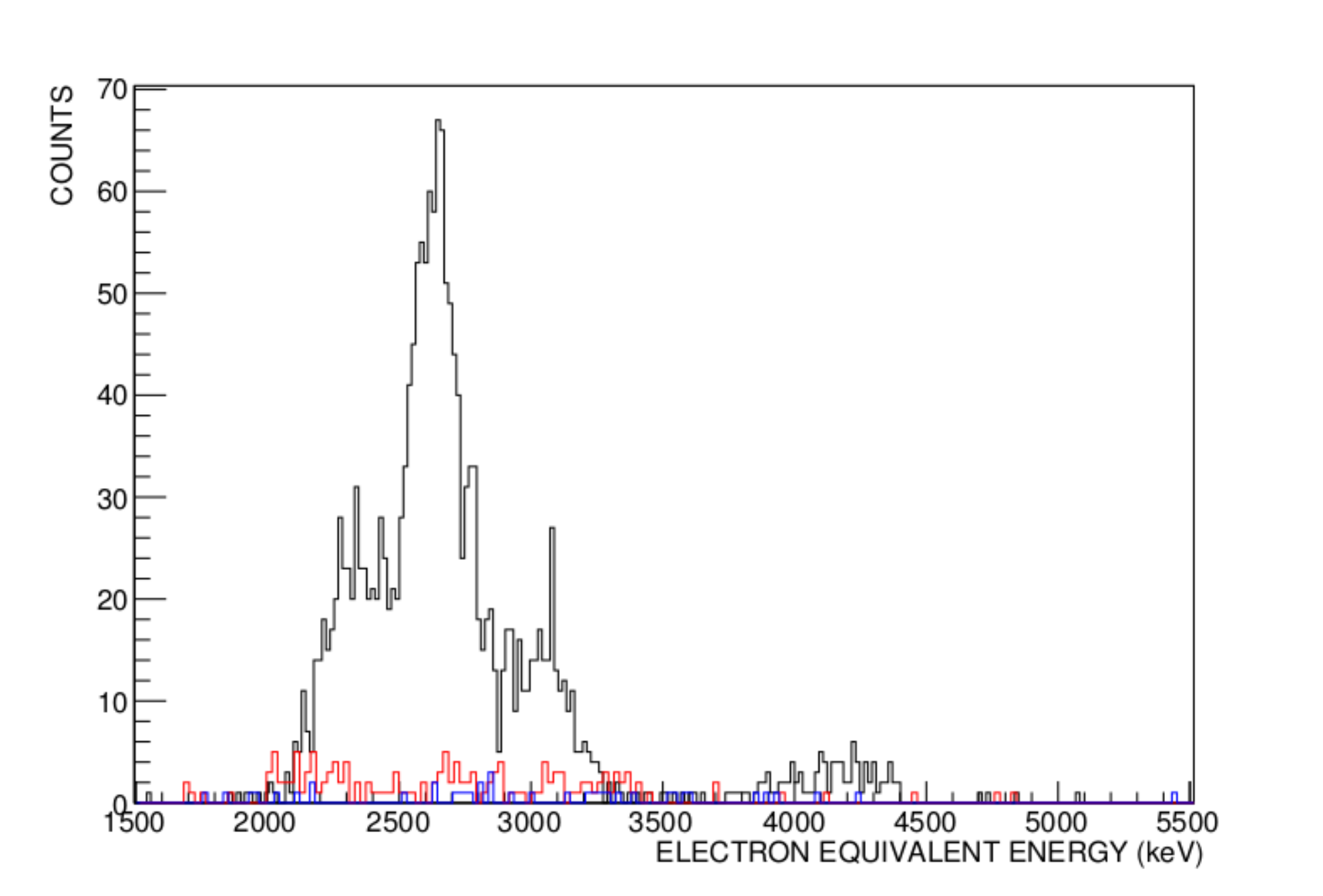}
\caption{The change of alpha ray intensity. Black: Ingot 16 made by normal crucible.
Red: Ingot 20 made by high purity graphite crucible. 
Blue: Ingot 24 made by further purified graphite crucible.}
\label{fg:change}
\end{wrapfigure}
The purification of NaI(Tl) crystal and selection of surrounding materials are present most important 
tasks to develop the PICO-LON detector system.
The RIs in NaI(Tl) crystal was effectively reduced by selecting surrounding materials which were 
used during purification and crystallization.
The environment around the laboratory was controlled by operating HEPA filter to reduce dust which 
attached the RIs.
The material of the graphite crucible was carefully selected by measuring the concentration in NaI(Tl) 
scintillator. 
The alpha ray intensities due to $^{238}$U, $^{226}$Ra, $^{232}$Th and their progeny are 
effectively reduced by selecting and purifying the material of the crucible as shown in 
Fig.\ref{fg:change}.
The intensity of the alpha ray due to  $^{210}$Po which is the progeny of $^{210}$Pb 
($T_{1/2}=22.4$ yr) was not reduced effectively by the purification of crucible.
We concluded that $^{210}$Pb was contaminated in raw powder of NaI\@.

After the selection of crucible was finished, further purification was done by chemical process of 
raw material of NaI powder.
The raw powder of NaI was solved into ultra pure water and it was poured into a column where 
cation exchange resin.
The resin was carefully selected to attach lead ions and radium ions effectively.
RIs in the resin were measured by using HPGe detector beforehand to avoid the contamination from 
the RIs in the resin.

The processed solution was dried by a rotary evaporator. 
The NaI powder was put into the high purity crucible and crystallization was done by 
hybrid Bridgemann method.
The parameters of crystallization were searched by several trial to make clear and without 
any inclusion crystals.

\section{Low background measurement}
The high purity NaI(Tl) crystal was cut to make a 7.62 cm$\phi\times$7.62 cm cylindrical shape.
One end and the side of the cylinder was covered with high purity white PTFE sheet to 
reflect scintillation photons and to guide the another end of the crystal.
A 4 mm thick high purity quartz light guide was glued on the end of the NaI(Tl) crystal.
The detector was contained in a high purity copper container to avoid deliquesce.

The low background measurement was done in the KamLAND area 
in Kamioka underground laboratory (36$^{\circ}$25'N, 137$^{\circ}$18'E) located at 2700 m 
water equivalent.
The experimental room was kept class 10 clean room by HEPA filter.
The NaI(Tl) detector was installed into a shield with 5 cm thick copper and at least 18 cm 
thck old lead.
Pure nitrogen gas was flushed near the NaI(Tl) detector to purge radioactive radon gas.
The exposure time was 26 days$\times$1.25 kg and the alpha ray energy spectrum was 
obtained by applying pulse shape discrimination analysis.

\begin{wrapfigure}{r}{0.45\textwidth}
\centering
\includegraphics[bb=0 0 777 661, width=\linewidth]{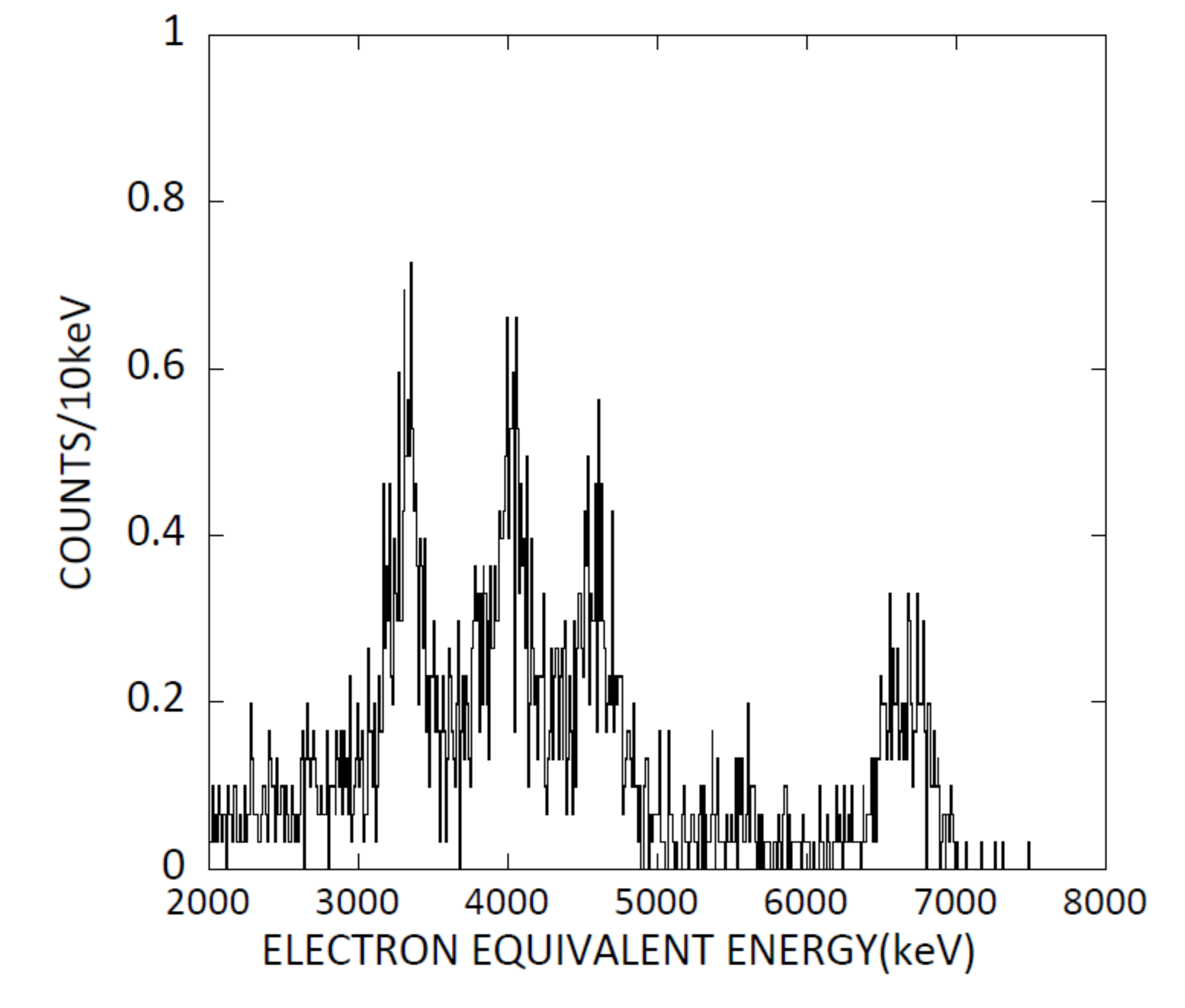}
\caption{Alpha ray spectrum taken by highly radiopure NaI(Tl) detector. }
\label{fg:Alpha}
\end{wrapfigure}

The obtained energy spectrum in the region of interest is shown in Fig.\ref{fg:Alpha}.
The energy of this spectrum is given by electron equivalent energy.
The electron equivalent energy is calibrated by the kinetic energy of electron which is 
produced by the interaction of gamma rays.
On the other hand, the scintillation output by heavy ions are less than the one by electrons.
The case of present NaI(Tl) scintillator, the quenching factor was 0.72.

The energy spectrum was fitted by several Gaussian function which corresponded to the 
U chain and Th chain alpha rays.
There were four prominent peaks at 3.4 MeV, 4.1 MeV, 4.6 MeV and 6.7 MeV\@.
These peaks are due to successive decays of $^{226}$Ra ($E_{\alpha}=4.784$ MeV), 
$^{222}$Rn ($E_{\alpha}=5.489$ MeV, $^{218}$Po ($E_{\alpha}=6.002$ MeV and 
$^{214}$Po ($E_{\alpha}=7.687$ MeV), respectively.
The detail of the analysis was described in the previous paper \cite{FushimiKEK}.
The fitted intensities of $^{226}$Ra, $^{222}$Rn and $^{218}$Po were the same because of 
the secular equilibrium of the radioactivity.
The intensity of $^{214}$Po was smaller than the other peaks because the half life of $^{214}$Po 
is 164 $\mu$sec. 
A significant amount of alpha ray events from $^{214}$Po is piled up with the beta ray 
which is emitted by $^{214}$Bi.

The results of U chain and Th chain reduction was listed in Table \ref{tb:purity}.
It is notable that the optimum selection of materials and chemical reduction are 
effective to purification of NaI(Tl) scintillator.
\begin{table}[thb]
\centering
\caption{Change of contamination of RIs in NaI(Tl) scintillator.
Concentration in this table is in unit of $\mu$Bq/kg.}
\label{tb:purity}
\begin{tabular}{l|lrrrr} \hline 
Ingot & Process & $^{238}$U & $^{226}$Ra & $^{210}$Pb & Th chain \\ \hline
16 &  Nomal crucible  & $520\pm73$ & $4510\pm60$ & $9600\pm100$ & $243\pm11$ \\
20 & Pure crucible & $372\pm23$ & $81\pm11$ & $440\pm22$ & $60\pm14$ \\ 
23\cite{FushimiKEK} & 20+Pb resin & $66\pm10$ & $108\pm18$ & $58\pm26$ & $13\pm8$ \\  
26\cite{FushimiTAUP2015} & 23+Ra resin & $  <0.5$  & $57\pm4$ & $29\pm7$ & $1.5\pm1.9$ \\ \hline
\end{tabular}
\end{table}

There is a remaining impurity of $^{40}$K which emits beta ray and 
gamma ray\cite{FushimiTAUP2015}.
The concentration of natural potassium in the ingot 26 was as large as 2.6 ppm, which was 
two orders of magnitude higher than the goal of the purity.
The chemical process to remove the potassium in NaI raw powder is now in progress.

\section{Future prospect}
We have successfully developed the highly radiopure NaI(Tl) crystal suitable to dark matter 
search with low RI impurities of $57\pm4$ $\mu$Bq/kg of $^{226}$Ra and $29\pm7$ $\mu$Bq/kg 
of $^{210}$Pb.
The present subjects to make larger detector are reduction of potassium concentration, 
development of a large volume ingot and selection of surrounding materials.
To develop a large volume ingot which enables to make 12.7 cm$\phi\times$12.7 cm 
crystal, we have selected the suitable material of crucible. 
The dimension of crucible was already determined and test crystallization was done.

The selection of surrounding material is the next important subject to develop high sensitivity
detector for dark matter.
All the materials which will be used for PMTs, housing, reflector are screened by measuring 
gamma rays from the samples.
The material selection is carried out by collaborating with XMASS group and support by 
Kavli IPMU funds (WPI) and 
Grant-in-Aid for Scientific Research on Innovative Areas number 26104008.
The background rate at the energy threshold ($\simeq 1$ keV electron equivalent) 
will be reduced to be less than 1 \dru \@.

A large volume NaI(Tl) array with 250 kg in total 
will be completed after five years and measurement of dark matter search will be started.

\section{Acknowledgment}
The authors thank Professor S.Nakayama for fruitful discussion and encouragement. 
The authors also thank Kamioka Mining and Smelting Company for supporting activities in the Kamioka mine and Horiba Ltd. for making the NaI(Tl) detectors. 
This work was supported by Grant-in-Aid for Scientific Research (B) number 24340055, 
Grant-in-Aid for Scientific Research on Innovative Areas number 26104008. 
The work was also supported by Creative Research Project in Institute of Socio, Arts and Sciences,
Tokushima University.

\end{document}